\documentclass[twocolumn,showpacs,preprintnumbers,amsmath,amssymb]{revtex4-1}

\usepackage{graphicx}
\usepackage{dcolumn}
\usepackage{bm}
\usepackage{mathrsfs}
\usepackage{soul}
\usepackage{color}
\setstcolor{blue}

\begin{document}

\preprint{APS/123-QED}

\title{McMillan-Rowell Oscillations in a Low Spin-Orbit SNS Semiconducting Junction }
\author{Binxin Wu$^{1}$, Chenxu Shao$^{1}$, Sherry Chu$^{1}$, B. Schmidt$^{1}$, M. Savard$^{1}$, Songrui Zhao$^{2}$, Zetian Mi$^{2}$, T. Szkopek$^{2}$, and G. Gervais$^{1}$}
\affiliation{$^{1}$Department of Physics, McGill University, Montreal, H3A 2T8 CANADA}
\affiliation{$^{2}$Department of Electrical and Computer Engineering, McGill University, Montreal, H3A 0C6 CANADA}

\date{\today}

\begin{abstract}
The electronic transport properties of an SNS junction formed by an InN nanowire (N) and Al contacts (S) with a superconducting transition temperature T$_c \simeq0.92$ K were investigated. As a function of \mbox{dc} bias, superconducting quasiparticle transport resonance peaks at $E=2\Delta$ were observed, in agreement with BCS theory with 2$\Delta(T=0) \equiv 2 \Delta_0$=275$\mu$eV. Several additional transport resonances scaling linearly in energy were observed at high-bias above 2$\Delta$, up to $E\simeq 15\Delta_0$, consistent with McMillan-Rowell oscillations. The persistence of McMillan-Rowell oscillations at high-bias and under applied magnetic field were investigated.

\end{abstract}

\pacs{74.45.+c,73.23.-b, 73.63.-b,}

\maketitle


There has been recent intense interest in the properties of hybrid structures composed of a low-dimensional semiconductor coupled by proximity effect to superconducting metal contacts\cite{YJDohDelft,RFrielinghaus,TNishioKoji,Spathis}. Following seminal theoretical work proposing the existence of engineered Majorana fermions in proximity-coupled strong spin-orbit semiconductor devices \cite{FuKane,JSauQW,JAlicea,JSauNW}, recent reports have shown encouraging experimental evidence for zero-energy Majorana modes\cite{VMourikDelft,ADasHeiblum,LRokhinson}. To observe Majorana modes, these studies have focused on large spin-orbit semiconductors such as InAs or InSb, and little attention has been devoted to similar devices fabricated with low spin-orbit material such as InN. We report here an extensive study of an SNS junction fabricated from an InN nanowire grown by molecular beam epitaxy (MBE). In addition to a zero-bias anomaly, non-linear transport resonances are clearly observed at bias energies up to fifteen times the superconducting gap $\Delta$. These resonances are consistent with McMillan-Rowell oscillations (MRO) \cite{Rowell} within the semiconducting nanowire. The persistence of MRO at high bias implies the persistence of Andreev reflection to high bias energy, suggesting that Andreev reflection at the semiconductor - superconductor interface is not fully understood.

The density of states, energy gap and Fermi velocity of a superconductor can be determined experimentally by tunnelling spectroscopy. This is usually achieved in an ideal SN junction by measuring the differential resistance or conductance as a function of \mbox{dc} bias energy $eV$ applied across the barrier. Generalized Andreev reflection in SNS tunnel junctions was described theoretically by Blonder, Tinkham, and Klapwijk (BTK)\cite{BTK,Tinkham} and has been successfully used to explain the tunnelling spectroscopy of junctions with a normal metal tunnelling barrier. However, in the case where a semiconductor takes the place of a normal metal, the band gap in the semiconductor density of states will modify the junction properties. In particular, the breakdown of the Andreev approximation $E >> E_F$ will modify the Andreev reflection probability at the interface. 

\begin{figure}
\includegraphics[width = 85mm]{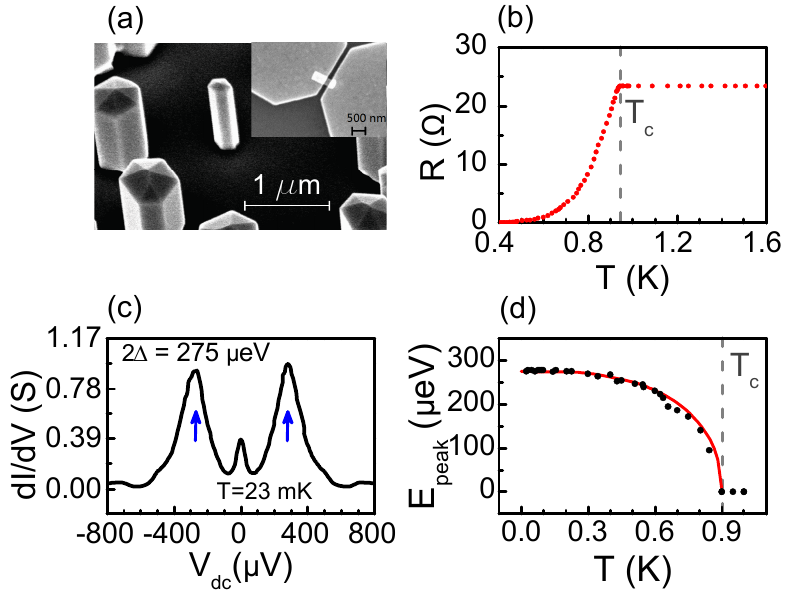}
\caption{\label{fig:RT} (a) TEM image of the as grown InN nanowires. In the inset, an SEM image of a representative SNS InN junction. The diameter of the studied nanowire is approximately 120~nm and its length $\sim$0.9~$\mu$m. (b) Temperature dependence of the resistance (at zero bias) in the temperature range 0.4~K to $\sim$ 1.6~K. The superconducting transition temperature is T$_c\simeq0.92$~K, indicated by the vertical dashed line.  (c) Differential conductance $dI/dV$ dependence on $V_{dc}$ at $T$~=~23~mK. The quasiparticle tunnelling peak is observed at $E_{peak}=2\Delta_0~=~275~\mu$eV.  (d) Temperature dependence of the quasiparticle peak position $E_{peak}$. The red solid curve is a fit of $2\Delta$ as a function of $T$ according to weak-coupling BCS theory \cite{Tinkham}.}
\end{figure}

We performed electron transport measurements on devices made with InN nanowires coupled to Al leads. The devices were fabricated from nearly intrinsic (not intentionally doped) non-tapered InN nanowires grown on Si (111) substrates by a Veeco Gen-II radio frequency plasma-assisted MBE growth system. Photoluminescent and XPS measurements show a band gap of approximately 0.675~eV and minimal surface charge effect on the nanowires \cite{Mi1,Mi2}. The wire  length is $\sim$ 0.9 ~$\mu$m and the diameter of the investigated nanowire is $\sim$120~nm. The $c$-axis is along the nanowire growth direction. To fabricate the ohmic contacts, the wires were transferred from the original (as-grown) substrate to an Si(100) wafer with a 100~nm thick SiO$_2$ insulating layer. Transfer was accomplished by diluting nanowires into an acetone solution in an ultrasonic bath, followed by drop casting the solution onto the target substrate. The substrate was pre-patterned with a mark array by photolithography to locate and identify each individual nanowire by scanning electron microscopy (SEM). Contact electrodes of Ti(10~nm)/Al(100~nm) were deposited by e-beam evaporation and lithography to establish superconducting contacts to the InN nanowires. An SEM image of a representative device is shown in  the inset of Fig.~\ref{fig:RT} (a), as well as the as-grown nanowires. Electrical transport measurements were performed in a dilution refrigerator equipped with a 9~T magnet down to temperatures of 23~mK. Magnetoresistance was measured by a constant-current source and a low-frequency \mbox{ac} and \mbox{dc} current bias technique. The differential resistance $dV/dI$ dependence on source-drain voltage$V_{dc}$ was obtained by a lock-in technique with an ac current of $I_{ac}$~=~50~nA, and the differential conductance $dI/dV$ was found by numerical inversion. In all data show in this work, a contact resistance $R_{contacts}=60\Omega$ was subtracted.

The contacts to the junction were verified to be ohmic. The two-point resistance was found to be less than 25~$\Omega$ in the normal state. Nanowires grown under the same conditions have been measured to give two-point resistances on the order of $\sim$1~M$\Omega$ in high-vacuum scanning probe experiments \cite{Mi3}. We attribute the discrepancy in resistance to electron doping of the nanowire. InN has a very high electron affinity ($\sim$5.8~eV \cite{Ager}), and will thus be electron doped by the Al contacts and also possibly by surface contaminants introduced during lithography and wet processing. The temperature dependence of the resistance was determined from 0.3 to~150~K, and is shown in Fig.\ref{fig:RT} (b) in the range between 0.4~K and $\sim$ 1.6~K. The abrupt drop in resistance at $T~\simeq$~0.92~K is consistent with previous work based on superconducting aluminum thin films. The coherence length of the Al contact $\xi= \sqrt{\hbar D/\Delta_0}\simeq 275 $ nm, assuming a diffusion coefficient for aluminum $D\simeq 160$ $cm^2 /s$, which is of order of the Al contact separation. From the Fermi velovity $v_{FN}\simeq 10^{5}$ $m/s$ in the nanowire, we estimate a temperature length $L_{T}=\hbar v_{FN}/k_{B}T\simeq 6$ $\mu m$ at 1K, suggesting that our junction is always in the mesocopic regime. 

The differential conductance measured by applying a mixed \mbox{dc} and \mbox{ac} current through the superconducting contacts is plotted in Fig.\ref{fig:RT} (c) at T~=~23~mK. Quasiparticle (QP) resonant peaks at $V_{dc}$~=~275~$\mu$eV are observed symmetrically in $dI/dV$ \textit{vs} $V_{dc}$ at both positive and negative biases. A conductance peak at zero bias is observed, whose origin is presently unknown and requires further investigation. Measurements of the $dI/dV$ spectra dependence on $V_{dc}$ were performed at several temperatures from 23~mK to $\sim$1~K.  The QP peak position $E_{peak}=eV_{dc}$ versus temperature is shown in Fig.\ref{fig:RT}(d), fitting well to a BCS theory of the gap\cite{Tinkham} indicated by the red line. The QP peak energy at $T=0$ can be inferred from the BCS fit, from which we find $2\Delta_0=275~\mu$eV. Our measurement is consistent with the expected BCS gap 2$\Delta_0^{BCS}=3.53\cdot k_B T_c =280~\mu$eV \cite{Tinkham} determined from the superconducting transition temperature $T_c=0.92$~K of Fig.~\ref{fig:RT} (b). The InN nanowire junction thus behaves consistently within weak-coupling BCS theory.

%
%
%
%

\begin{figure}[hbt]
\includegraphics[width = 80mm]{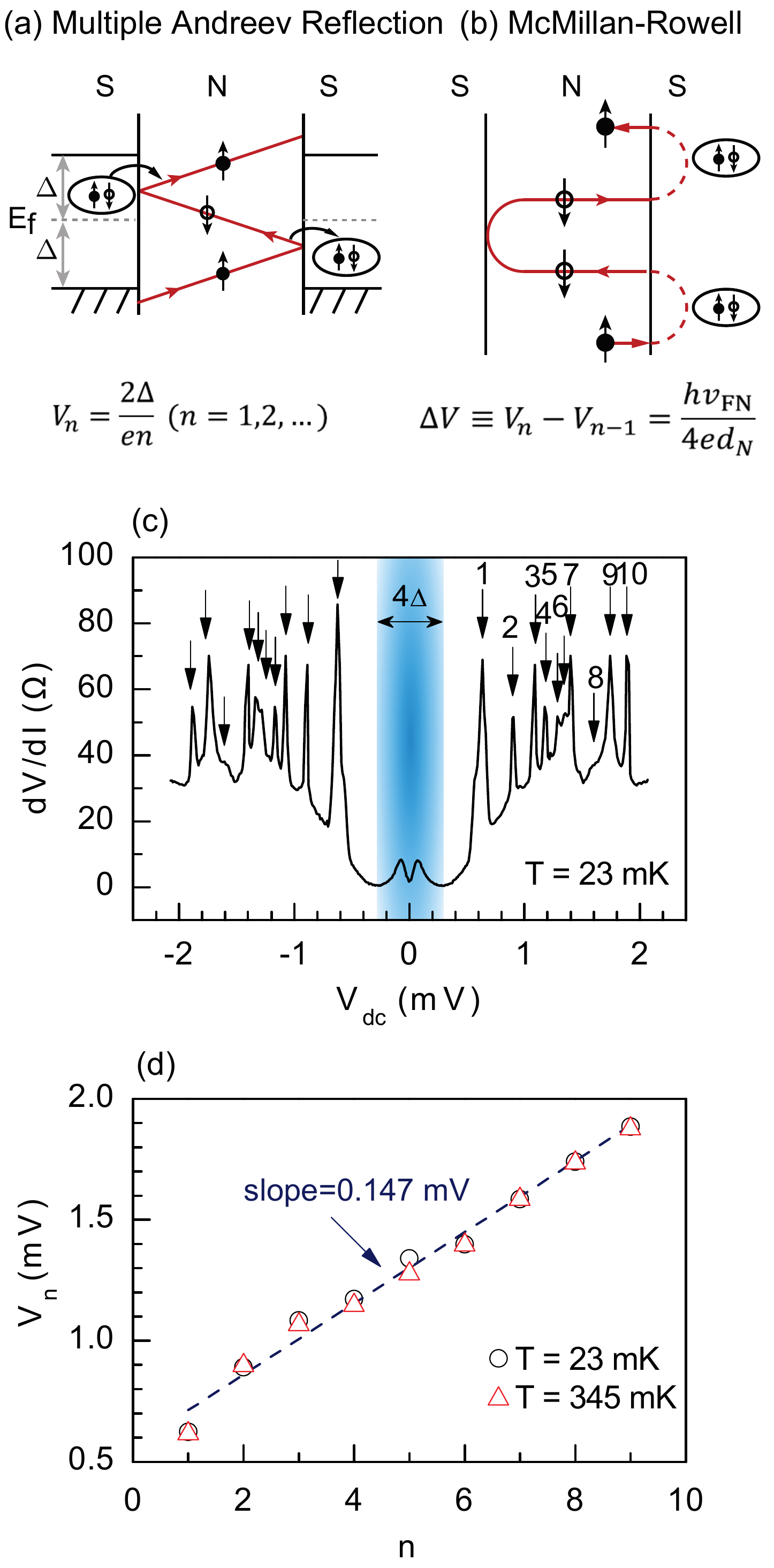}
\caption{ (a) Schematic of multiple Andreev reflection ($n=1$). Interference occurs due to Andreev reflection at each interface. (b) Schematic of McMillan-Rowell oscillations ($n=1$). Interference occurs due to Andreev reflection and normal reflection. (c) $dV/dI$ \textit{vs} $V_{dc}$ spectrum at $T=23$~mK. At bias above the gap $eV_{dc}$$>$2$\Delta_0$, transport resonances labeled by order $n$ are observed at voltages $V_{n}$. (d) $V_n$ \textit{vs} $n$ at $T=23$~mK (black circles) and $345$~mK (red triangles). A linear fit is indicated by the dotted line. }
\label{fig:dIdVpeak}
\end{figure}

In addition to superconducting QP peaks at 2$\Delta$~=~0.275~meV, numerous other transport resonances are observed in $dV / d I$ at bias energies above 2$\Delta$ as shown in Fig.~\ref{fig:dIdVpeak}(c). The $dV/dI$ resonances beyond 2$\Delta$ are enumerated by  $n$, and the resonant bias voltage is plotted versus $n$ in Fig.~\ref{fig:dIdVpeak}(d) at 23~mK and 345~mK. A clear linear increase in resonant bias  is observed with a slope 0.147$\pm$0.007~mV, with a noticeable deviation occurring at $n = 1 \rightarrow 2$. Notably, as the temperature is increased the $dV/dI$ resonances weaken in amplitude and eventually disappear as $T\rightarrow T_c$. This suggests that these features are directly related to a transport mechanism directly tied to the superconducting gap, implying important contributions from Andreev reflections.

We now turn to an interpretation of the transport resonances above the BCS gap. At temperatures below the superconducting transition temperature $T_c$, transport resonances can occur due to Andreev reflection at SN interfaces. Multiple Andreev reflection (MAR) shown schematically in Fig.~\ref{fig:dIdVpeak}(a) results in transport resonances at sub-gap energies $V_n=2\Delta/en$, where $n=1,2,3,...$ is the interference order. MAR is expected to be strong in SNS junctions where the transparency of each SN interface is similar, and greatly suppressed in junctions with significant SN interface asymmetry. Sub-gap transport resonances are notably absent in the InN nanowire junction, as seen in Fig.~\ref{fig:dIdVpeak}(c). On the other hand, MRO \cite{ONesher,Rowell,WJGallagher,Ingerman} are transport resonances that can occur in an asymmetric SNS junction, resulting from normal reflections at the opaque SN interface and Andreev reflection at the transparent SN interface, depicted in Fig.~\ref{fig:dIdVpeak}(b). The reflections establish a series of geometrical resonances leading to MRO, and were first observed by direct measurement of density of states. MRO also  manifests itself in a series of equidistant peaks in the dynamic transport spectrum,
\begin{eqnarray}
\Delta V\equiv V_n-V_{n-1}=\frac{hv_{FN}}{4ed_N},
\label{eq:two}
\end{eqnarray}
where $v_{FN}$ is the Fermi velocity in the normal material, and $d_N$ is the effective length of the normal region of the junction. Note that contrary to the typical situation of an SNS junction with a metallic normal region, the Fermi velocity of a semiconducting normal region is tuneable with carrier density, and the Fermi energy much smaller. As shown in Fig.~\ref{fig:dIdVpeak}(b), the transport resonances in our SNS junction are observed to be equidistant, scaling linearly with $n$ with a slight deviation at $n=1$, in good agreement with MRO. The clear observation of these sharp and
well-defined MROs suggest a ballistic transport occurring in the nanowire. The slope of $V_n$ versus $n$ and the nanowire length $d=0.9\mu\mathrm{m}$ together with Eq.~\ref{eq:two} allows us to estimate an electron Fermi velocity in the InN nanowire of $v_{FN}=1.3\times10^7~$cm/s.
Furthermore, assuming a three-dimensional density of states for the relatively large diameter nanowire we infer an electron density of order of $n\simeq 1.6\times 10^{16}$ $cm^{-3}$. This density agrees within an order of magnitude with previous measurements using an electrical nanoprobing technique\cite{Mi3}. Taking the effective mass of InN as $m^*=0.07~m_e$, the corresponding Fermi energy is estimated to be $E_{FN}\simeq 3.3~\mathrm{meV}$ above the conduction band edge. The ratio of Fermi energy to the superconducting gap is $E_{FN} / \Delta_0 \simeq 20$, which is several orders of magnitude smaller than can be typically achieved in an SNS junction with a metallic normal region. Importantly, BTK theory of the SNS junction is valid in the limit $E_{FN}/\Delta_0 \rightarrow \infty$, a valid approximation for metallic normal regions but which is expected to break down for junctions with semiconducting normal regions of low carrier density. A surprising feature of the transport resonances is the persistence in resonance amplitude up to an energy of $15\Delta_0$, in contradiction with the expected rapid collapse of Andreev reflection probability (less than $10^{-2}$) at large bias in SNS junctions with metallic normal regions \cite{BTK}. The persistence of MRO to high bias has however  been experimentally observed in previous work with cuprates, up to energies ten times the superconducting gap  \cite{ONesher}.

\begin{figure}[hbt]
\includegraphics[width = 80mm]{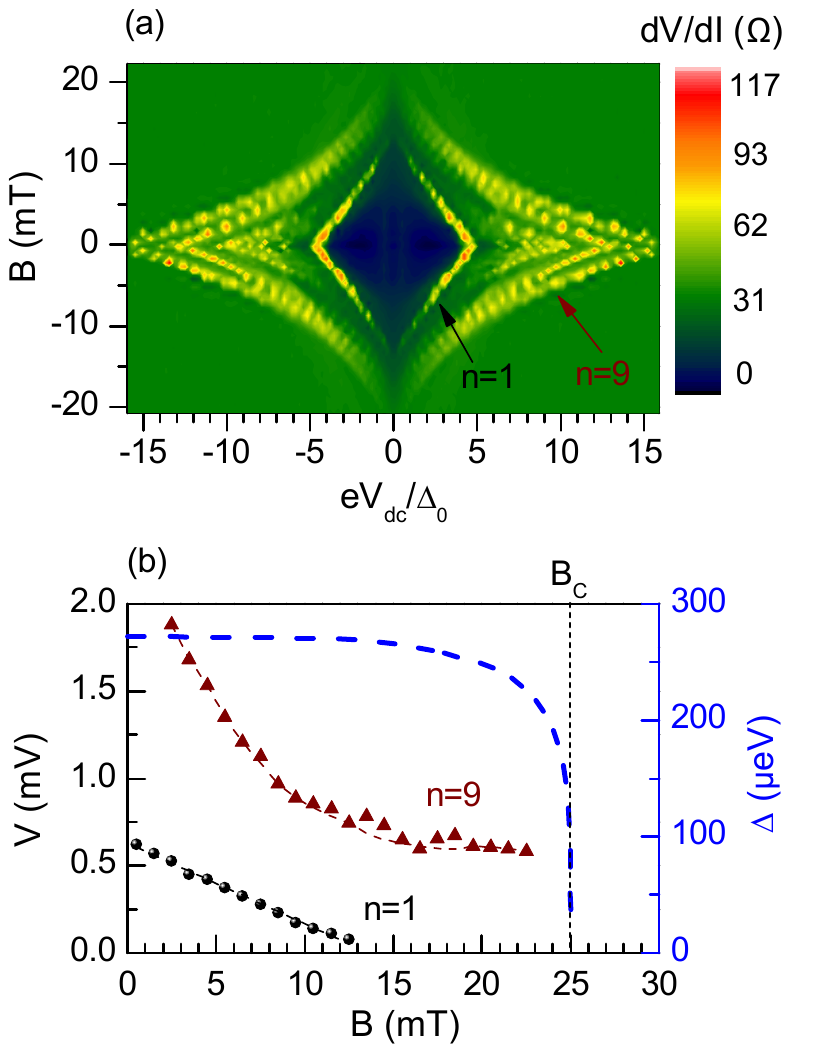}
\caption{(a) Contour plot of $dV/dI$ $vs$ $eV_{dc}$/$\Delta_0$ and applied magnetic field $B$ (open symbols) at a temperature of 23~mK. (b) Dispersion of MRO peaks $V_{1}$ and $V_{9}$ versus $B$ (filled symbols). A BCS model for the $B$-field dependence of $\Delta(B)$ is also plotted (dotted line), showing a different dispersion than MRO.}
\label{fig:dVdIB}
\end{figure}

The persistence of MRO was further investigated in the presence of a magnetic field applied perpendicular to the substrate and nanowire. The measured $dV/dI$ is shown in Fig.~\ref{fig:dVdIB} (a) as a two-dimensional contour plot of $dV/dI$ versus $eV_{dc}$/$\Delta_0$ over magnetic field $B$ ranging from 22.5~mT to -20.5~mT.  The central dark diamond shaped region of low resistivity includes the QP peaks in $dI/dV$ (dips in $dV/dI$). The QP resonances disappear with applied magnetic field, as expected from the weakening of the BCS gap under magnetic field. The MRO peaks in $dV/dI$, indicated by bright lines, evolve with magnetic field and eventually disappear at a critical field, thereby recovering the fully linear character of electron transport. The evolution of the first and last observed MRO peak, $V_1$ and $V_{9}$, is plotted versus $B$ in Fig.~\ref{fig:dVdIB} (b). For comparison, the scaling of the BCS gap with magnetic field is also illustrated by a dotted line. Although all features are seen to disappear at a critical field, the dispersion of $V_1$ and $V_9$ versus $B$ differ with each other and with the trend of the weak-coupled BCS gap. The dispersion in threshold may arise from a variety of magnetic field induced effects, such as the modulation of the coherence length $\xi{(B)}$, modulation of Andreev reflection probability, and magnetic confinement. Further experimental and theoretical work is required to better understand the magnetic field behaviour of low spin-orbit InN SNS junctions.

In conclusion, we have investigated the transport properties of a hybrid SNS device formed with a low spin-orbit InN nanowire. The sharp, non-linear transport resonances observed at energies well above the superconducting gap are attributed to MRO, and are amongst the most pronounced observations of MRO in electrical transport. The persistence of MRO to biases well above gap suggest that Andreev reflection at interfaces of semiconductors and conventional s-wave superconductors persist to high energies unlike the collapse of Andreev reflection probability at high bias at an interface of normal metal and semiconductor. The observed persistence and dispersion of MRO, and thus Andreev reflection, in an SNS junction with semiconducting normal region is presently not understood. In order to realize devices from heterostructures composed of semiconductor / superconductor junctions, further work is required to understand the nature of basic electron transport phenomena. In particular, an extension of BTK theory to account for situations of small Fermi energy may be required to describe Andreev processes in weakly doped semiconductors.

We thank S. Bohloul, H. Guo and A. Clerk for stimulating discussions. We acknowledge the technical assistance of Richard Talbot, Robert Gagnon and John Smeros. This work was supported by NSERC (Canada), FRQNT (Qu\'ebec), and CIFAR. We also thank the McGill Nanotools Microfabrication Facility.

\appendix

\end{document}